\documentclass[10pt,twocolumn,letterpaper，threeparttable]{article}

\usepackage{wacv}
\usepackage{times}
\usepackage{color}
\usepackage{epsfig}
\usepackage{graphicx}
\usepackage{amsmath}
\usepackage{amssymb}
\usepackage{subfigure}
\usepackage{adjustbox}
\usepackage{algorithm}
\usepackage{algpseudocode}
\usepackage{booktabs} 

\usepackage[pagebackref=true,breaklinks=true,colorlinks,bookmarks=false]{hyperref}

\wacvfinalcopy 

\def\wacvPaperID{961} 

\ifwacvfinal\fi
\begin{document}

\title{A Self-attention Residual Convolutional Neural Network for Health Condition Classification of Cow Teat Images}

\author{Minghao Wang \  \\
Yeshiva University\\
{\tt\small mwang3@mail.yu.edu}
}

\maketitle
\ifwacvfinal\thispagestyle{empty}\fi

\begin{abstract}

Milk is a highly important consumer for Americans and the health of the cows' teats directly affects the quality of the milk. Traditionally, veterinarians manually assessed teat health by visually inspecting teat-end hyperkeratosis during the milking process which is limited in time, usually only tens of seconds, and weakens the accuracy of the health assessment of cows' teats. Convolutional neural networks (CNNs) have been used for cows' teat-end health assessment. However, there are challenges in using CNNs for cows' teat-end health assessment, such as complex environments, changing positions and postures of cows' teats, and difficulty in identifying cows' teats from images. To address these challenges, this paper proposes a cows' teats self-attention residual convolutional neural network (CTSAR-CNN) model that combines residual connectivity and self-attention mechanisms to assist commercial farms in the health assessment of cows' teats by classifying the magnitude of teat-end hyperkeratosis using digital images. The results showed that upon integrating residual connectivity and self-attention mechanisms, the accuracy of CTSAR-CNN has been improved. This research illustrates that CTSAR-CNN can be more adaptable and speedy to assist veterinarians in assessing the health of cows' teats and ultimately benefit the dairy industry.

\end{abstract}

\section{Introduction}

Milk consumption holds significant importance in American society, and the United States proudly stands as the largest producer of milk worldwide. Renowned for its abundant nutritional content, milk is often considered a dietary necessity. The quality of milk is directly influenced by the health of the cows producing it. Among the various health issues affecting cows, mastitis stands out as one of the most critical problems~\cite{NF2001}. Detecting the risk of mastitis is crucial for ensuring the overall well-being and productivity of dairy cows. Frequent monitoring of teat-end callosity is critical for a mastitis prevention program~\cite{WD1982}. Traditionally, veterinarians have relied on assessing the health of a cow's teat during the milking process. However, this assessment is typically time-limited, spanning only tens of seconds, which poses limitations on the accuracy of mastitis risk assessment.
In commercial dairy farms, cows are typically moved to the milking parlor where their milking sessions are recorded by cameras. Leveraging the advancements in computer vision and machine learning techniques, researchers have explored the utilization of recorded cow teat videos to identify cows's teats~\cite{Zhang2022}. veterinarians gain the flexibility to conduct more comprehensive health assessments of the teat. This approach offers the potential to enhance the efficiency and accuracy of mastitis risk assessment in dairy cows. Furthermore, deep learning (DL) has been proposed to classify the extent of hyperkeratosis using a four-level classification scheme~\cite{PI2021} and the accuracy of this approach was 46.7–61.8\%.

Several challenges arise when utilizing images of cows' teats for mastitis risk assessment. Firstly, the environment within the milking parlor can be complex, potentially resulting in issues such as insufficient lighting or obstructed views of the camera. These factors may adversely affect the quality and usability of the data, impeding accurate analysis. Secondly, the position and posture of the cows' teats in the images can vary, which complicates the identification process. Lastly, effective health assessment of cows' teats requires consideration of multiple factors, including teat size, shape, and color. The intricacies associated with these factors compound the challenge of accurate mastitis risk assessment within the image footage.

\begin{figure}[h]
\centering
\includegraphics[width=6cm]{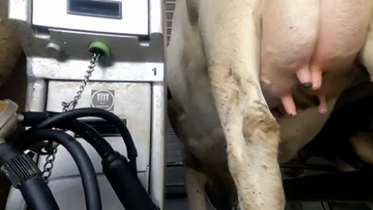}
\caption{In commercial dairy farms, a cow is moved to the milking parlor and recorded by cameras.}
\label{fig:Regression-net}
\end{figure}

To address the aforementioned challenges and improve the performance of health assessment models for cows' teats, a cows' teats self-attention residual convolutional neural network (CTSAR-CNN) model is proposed. The performance of the original CNN model is improved via a combination of residual connectivity and self-attention mechanisms. This paper provides two specific contributions: 1) CTSAR-CNN is proposed to improve accuracy for the cows' teats image classification; 2) The effectiveness of residual convolutional and self-attention mechanisms for performance improvement of CNN model in cows' teats image classification is verified.

\section{Related Work}\label{sec:related}
\subsection{Neural Network for Images Classification}
Neural network models have a long history of development. In the last few decades, neural networks have moved from the laboratory into the industry. The development of neural network models began in 1943 and was inspired by the layout of biological neurons\cite{mcculloch1943logical}. These neural networks were built in layers, with the input of one layer connected to the output of the next. At that time mathematician Walter Pitts and neurophysiologist Warren McCulloch researched and wrote a paper describing how neurons might work and developed a simple neural network using electrical circuits\cite{piccinini2004first}. 

By the 1980s, artificial neural networks (ANN) had become a hot topic in the field of artificial intelligence. Researchers realized that adding just a few hidden layers could greatly enhance the capabilities of their neural networks, leading to the development of ANN\cite{sejnowski2020unreasonable}. At the same time, advances in materials science, particularly the development of metal oxide semiconductor (MOS) very large scale integration (VLSI) in the form of complementary MOS (CMOS) technology, enhanced the performance of computers and enabled ANNs to move from theory to implementation. 

Convolutional neural network (CNN) is a kind of neural network model commonly used in the field of computer vision, and its origin can be traced back to the "neocognitron" introduced by Kunihiko Fukushima in 1980\cite{Fukushima1980}. In the 21st century, CNN has developed rapidly. With advances in deep learning theory and improvements in numerical computing devices, CNN has made great progress in computer vision, natural language processing, and other fields\cite{Ferraz2021}. Deep learning, a subset of machine learning that uses neural networks with multiple hidden layers, has become the dominant approach in many fields, including computer vision, natural language processing, and speech recognition. By now, there have been many advances in the field of neural networks, including the development of new architectures, training algorithms, and applications. The use of neural networks has also spread beyond academia into industry, with many companies using them to develop new products and services.

Neural network models have developed rapidly in recent years. The field of computer vision has also made great progress with the introduction of several models. The ResNet (residual network) model was proposed by Kaiming He et al. in their paper “Deep Residual Learning for Image Recognition”\cite{He2015}. The ResNet model introduces the concept of residual learning, which is the use of fast connections to skip one or more layers. This allows the neural network to learn the residual function and thus improve the performance of image classification tasks. The Inception model was proposed by Christian Szegedy et al. in their paper “Going Deeper with Convolutions”\cite{Szegedy2015}. Inception introduces the concept of Inception modules, which are designed to efficiently extract features at multiple scales. The Xception model was proposed by François Chollet in his paper "Xception: Deep Learning with Depthwise Separable Convolutions" \cite{Chollet2017}. As the name implies, the Xception architecture performs slightly better than Inception V3 on the ImageNet dataset and significantly better than it on a larger image classification dataset. on a larger image classification dataset. The ViT (Vision Transformer) model was introduced by Alexey Dosovitskiy et al. in their paper “An Image is Worth 16x16 Words: Transformers for Image Recognition at Scale”\cite{Dosovitskiy2020}. The ViT model applies the transformer architecture originally designed for text-based tasks to image recognition. the ViT model represents the input image as a series of image patches and directly predicts the category labels of the image.

Neural network models have been used to classify images. Image classification is a computer vision task that automatically assigns images to one or more predefined categories or labels. For example, an image classification system may recognize the presence of cats, dogs, vehicles, etc. in a picture. The process typically involves feature extraction (extracting useful information from the image), feature selection (determining which features are most important for classification), and classification (using algorithms such as support vector machines, neural networks, etc. to classify the features). Image classification is important in a variety of applications such as facial recognition, medical image analysis, vehicle detection, etc.

\subsection{Neural Network for Health Assessment of Cows' Teats}
In the realm of dairy production, the pivotal role played by the health of dairy cows is unequivocal, particularly in influencing the quality and yield of milk. A critical health concern that significantly affects milk-producing bovines is mastitis, a condition predominantly impacting the cows' teats. The assessment and management of this condition are central to ensuring optimal dairy production.

\vspace{-1pt} 
The prevalent method for evaluating the risk of mastitis involves examining the condition of the cow's teat, specifically through the lens of teat hyperkeratosis. This condition is typically categorized using a four-tier scoring system, which classifies the state of the teat-end based on varying degrees of hyperkeratosis. The classifications range from Score 1, indicating the absence of a ring, to Score 4, which signifies a very rough ring\cite{MG2001}. This systematic classification plays a vital role in standardizing the assessment of teat health.

\vspace{-1pt} 
Traditionally, the task of evaluating teat health has been primarily undertaken by veterinarians during the milking process. While this approach provides direct observation, it is fraught with challenges such as time consumption, labor intensity, and potential subjectivity, leading to a lack of precision in assessments. To address these issues, there has been a burgeoning interest in leveraging neural networks for a more accurate and efficient assessment of cows' teat health.

\vspace{-1pt} 
Numerous studies have underscored the efficacy of neural networks in this domain. For instance, Y Zhang et al. introduced an innovative unsupervised, sample-less method for extracting keyframes from video footage of the milking process, a technique that has shown promise in practical applications on commercial farms\cite{Zhang2022}. Similarly, M Wang et al. developed this approach by integrating fusion deep distance and ensemble modeling techniques to identify keyframes in cow teat videos\cite{MW2023}. Furthermore, I.R. Porter et al. used the GoogLeNet architecture to accurately score the degree of teat hyperkeratosis using the established 4-point scale\cite{PI2021}. In another notable advancement, Y Zhang et al. proposed a separable confident transductive learning (SCTL) model, which has demonstrated significant improvements in the performance of teat-end image classification\cite{zhangdairy2022}\cite{zhang2022separable}.

\vspace{-1pt} 
The utilization of neural networks in the tasks of identification, segmentation, and classification of cow teats represents a significant technological leap in this field. By enabling more precise and efficient health assessments of cows' teats, these advancements not only contribute to the health and well-being of the animals but also have a cascading positive impact on the productivity and sustainability of the dairy industry. This paradigm shift towards the adoption of neural network-based techniques marks a pivotal step in modernizing dairy farm practices, aligning them with contemporary technological advancements, and ensuring a more robust and resilient dairy production chain.

\section{Methods}\label{sec:method}

\subsection{Motivation}
There are already images that have been classified. Based on the uniqueness of the data and the problem, supervised learning is used to classify images. There is designed the following process for image classification from cows' teats images using classified images as training samples with images directly resized to 224x224 pixels.

\begin{figure}[h]
\centering
\includegraphics[width=6cm]{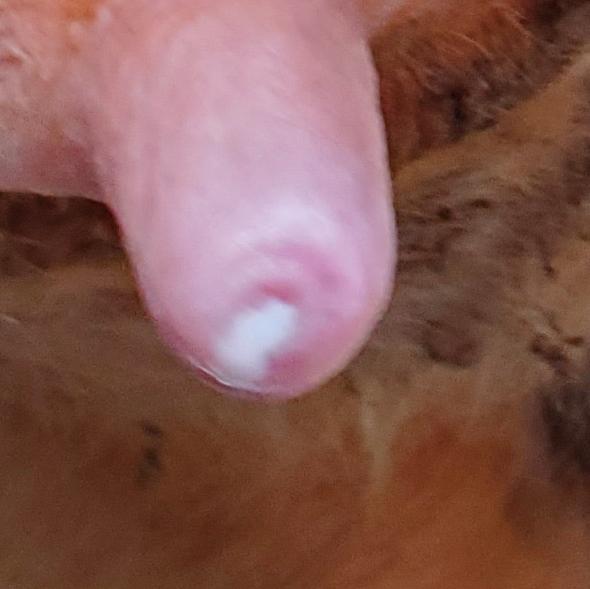}
\caption{Sample Original Image.}
\label{fig:Regression-net}
\end{figure}

\begin{table}[h]
\centering
\caption{Class Distribution}
\begin{tabular}{|l|c|r|}
\hline
class   & count & percentage \\
\hline
score1  & 450   & 39.16\%    \\
score2  & 491   & 42.73\%    \\
score3  & 187   & 16.28\%    \\
score4  & 21    & 1.83\%     \\
\hline
total   & 1149  & 100\%      \\
\hline
\end{tabular}
\end{table}

\subsection{CTSAR-CNN Model}
The CTSAR-CNN Model model has five convolutional layers, five pooling layers, five resblocks, and 2 self-attention. The structure of the model is in table 2, where K, S, and P denote kernel, stride, and padding, respectively.

Table 2 titled "CTSAR-CNN Model Architecture" provides a detailed layout of a convolutional neural network specifically tailored for the classification of cow teat health. It enumerates a sequence of layers, their types, the transformation of feature maps, and the configuration parameters for each layer. The layers are listed sequentially from 1 to 21, with certain numbers omitted for brevity. The types of layers include convolutional layers (Conv2d), residual blocks (ResBlock), max-pooling layers (MaxPool2d), self-attention mechanisms (SelfAttention), adaptive average pooling layers (AdaptiveAvgPool2d), and fully connected layers (Linear). The transition of feature map sizes is denoted by arrows, indicating the depth of the feature maps at each layer. Parameters for each layer are specified in terms of kernel size (K), stride (S), and padding (P), with dashes indicating default or inapplicable settings for certain layers. The architecture culminates in a fully connected layer that maps to four classes, reflecting the number of classes in the classification task. This structured approach illustrates the complexity and depth of the model designed to process and analyze image data for the targeted application.

\begin{table}[h]
\centering
\caption{CTSAR-CNN Model Architecture}
\begin{adjustbox}{margin=-0.7cm 0cm 0cm 0cm}
\begin{tabular}{|c|c|c|c|}
\hline
Layer & Type & Feature Maps & K/S/P \\
\hline
1 & Conv2d & 3 $\rightarrow$ 64 & 3x3 / - / 1 \\
2 & ResBlock & 64 & - \\
3 & MaxPool2d & - & 2x2 / 2 / - \\
4 & SelfAttention & 64 & - \\
\hline
5 & Conv2d & 64 $\rightarrow$ 128 & 3x3 / - / 1 \\
6 & ResBlock & 128 & - \\
7 & MaxPool2d & - & 2x2 / 2 / - \\
\hline
8 & Conv2d & 128 $\rightarrow$ 256 & 3x3 / - / 1 \\
9 & ResBlock & 256 & - \\
10 & MaxPool2d & - & 2x2 / 2 / - \\
\hline
12 & Conv2d & 256 $\rightarrow$ 512 & 3x3 / - / 1 \\
13 & ResBlock & 512 & - \\
14 & MaxPool2d & - & 2x2 / 2 / - \\
15 & SelfAttention & 512 & - \\
\hline
16 & Conv2d & 512 $\rightarrow$ 512 & 3x3 / - / 1 \\
17 & ResBlock & 512 & - \\
18 & MaxPool2d & - & 2x2 / 2 / - \\
\hline
19 & AdaptiveAvgPool2d & - & 3x3 \\
20 & Linear & 512 * 3 * 3 $\rightarrow$ 1024 & - \\
21 & Linear & 1024 $\rightarrow$ 4 & - \\
\hline
\end{tabular}
\end{adjustbox}
\end{table}

\subsection{Model Training and Hyperparameters}
The training process employs the Adam optimizer and uses cross-entropy as the loss function. The learning rate is set to 0.001, and the weight decay is also set to 5e-4. Batch size is 32. Weights were assigned to different kinds of images based on their quantitative differences. Select 15\% of the images in the training set as the validation set. Accuracy evaluation was performed using Cow\_teat\_classfication\_accuracy\cite{zhang2022separable} \cite{zhangdairy2022}.

\section{Results}\label{sec:results}

\begin{table}[h]
\centering
\caption{Result}
\begin{tabular}{|c|c|}
\hline
\textbf{Model} & \textbf{Accuracy} \\
\hline
GoogLeNet & 61.8\% \\
\hline
CTSAR-CNN & 62.6\% \\
\hline
\end{tabular}
\end{table}

The model was trained for 50 epochs, and on the training set, the loss stabilized at 0.9 and the accuracy stabilized at about 61\%; on the validation set, the loss stabilized at 1.0, and the accuracy also stabilized at about 61\%. The final accuracy of CTSAR-CNN on the test set is 62.63\%. The results show that CTSAR-CNN has better performance compared to GoogLeNet.

\section{Discussion}\label{sec:dis}

The CTSAR-CNN, a convolutional neural network, has been in advancing the classification of cows' teat images. Although this model marks a progressive step beyond the capabilities of GoogLeNet, its improvements in accuracy and efficiency remain modest. A critical analysis suggests that one of the primary factors contributing to this limited enhancement is the dataset itself, particularly the constrained volume of training data available. The quantity and diversity of data play a pivotal role in the efficacy of machine learning models, as they directly impact the model's ability to learn and generalize from various examples. Therefore, the current limitation in the dataset size and variety could be a significant bottleneck, hindering the model's potential to reach its full analytical capacity.

Addressing this challenge necessitates a two-pronged strategy centered on expanding the dataset and employing sophisticated digital image augmentation techniques. Firstly, a concerted effort to augment the dataset is crucial. This can be achieved by collecting a more extensive and diverse array of cows' teat images, thereby enriching the training set with a wider range of examples. Such an expansion would not only enhance the depth and breadth of the dataset but also provide a more comprehensive representation of the variations found in real-world scenarios. Secondly, the application of digital image enhancement techniques is vital. By incorporating methods such as random rotations, crops, and adjustments in brightness, saturation, and contrast, the existing data can be manipulated to simulate a more varied set of training examples. This process of image augmentation serves to artificially expand the dataset, introducing a level of variation and complexity that could aid in training more robust models. Additionally, fine-tuning the model's learning parameters, such as experimenting with different learning rates, applying various regularization techniques, and employing early stopping criteria, could further optimize the model's performance. These strategies, collectively, hold the potential to significantly elevate the accuracy and reliability of the CTSAR-CNN in classifying cows' teat images, ultimately leading to more effective and precise applications in the dairy industry.

\section{Conclusion}\label{sec:conclusion}
This paper proposed the CTSAR-CNN model, a novel approach to assist the health assessment of cows' teats using self-attention and residual connectivity. CTSAR-CNN model outperformed the existing GoogLeNet model, demonstrating modest improvements with the incorporation of these advanced mechanisms. The research emphasizes the potential of using deep learning for veterinary diagnostic tasks, showing that accurate, non-invasive assessments can aid in early detection and treatment, ultimately enhancing dairy cow productivity. Future work will aim to expand the dataset and explore more sophisticated image augmentation techniques to further improve the model's performance and robustness. Ongoing research efforts will focus on refining the model's ability to generalize across different breeds and environmental conditions, ensuring it can be reliably deployed in diverse farm settings.



{\small
\bibliographystyle{unsrt}
\bibliography{egbib}

\begin{thebibliography}{10}

\bibitem{NF2001}
F~Neijenhuis, H~W Barkema, H~Hogeveen, and J~P Noordhuizen.
\newblock Relationship between teat-end callosity and occurrence of clinical mastitis.
\newblock {\em J. Dairy Sci.}, 84:2664--2672, 2001.

\bibitem{WD1982}
D~Williams and G~Mein.
\newblock Physical and physiological factors affecting milk flowrate from the bovine teat during machine milking.
\newblock {\em Occas. Publ. N. Z. Soc. Anim. Prod. (N. Z.)}, 8:42--74, 1982.

\bibitem{Zhang2022}
Y~Zhang, M~Wieland, and P~Basran.
\newblock Unsupervised few shot key frame extraction for cow teat videos.
\newblock {\em Data}, 7(5):68, 2022.

\bibitem{PI2021}
I~Porter, M~Wieland, and P~Basran.
\newblock Feasibility of the use of deep learning classification of teat-end condition in holstein cattle.
\newblock {\em J. Dairy Sci.}, 104:4529--4536, 2021.

\bibitem{mcculloch1943logical}
W~Pitts W~McCulloch.
\newblock A logical calculus of the ideas immanent in nervous activity.
\newblock {\em Bulletin of Mathematical Biophysics}, 5:115--133, 1943.

\bibitem{piccinini2004first}
G~Piccinini.
\newblock The first computational theory of mind and brain: A close look at mcculloch and pitts's logical calculus of ideas immanent in nervous activity.
\newblock {\em Synthese}, 141(2):175--215, 2004.

\bibitem{sejnowski2020unreasonable}
TJ~Sejnowski.
\newblock The unreasonable effectiveness of deep learning in artificial intelligence.
\newblock {\em PNAS}, 117(48):30033--30038, 2020.

\bibitem{Fukushima1980}
K~Fukushima.
\newblock Neocognitron: A self-organizing neural network model for a mechanism of pattern recognition unaffected by shift in position.
\newblock {\em Biol. Cybern.}, 36(4):193--202, 1980.

\bibitem{Ferraz2021}
C~Ferraz, C~Toledo, and et~al.
\newblock An introduction of cnn: Models and training on neural network models for image classification.
\newblock In {\em IEEE}, 2021.

\bibitem{He2015}
K~He, X~Zhang, S~Ren, and J~Sun.
\newblock Deep residual learning for image recognition.
\newblock {\em arXiv preprint arXiv:1512.03385}, 2015.

\bibitem{Szegedy2015}
C~Szegedy, W~Liu, Y~Jia, P~Sermanet, S~Reed, D~Anguelov, and et~al.
\newblock Going deeper with convolutions.
\newblock In {\em Proceedings of the IEEE conference on computer vision and pattern recognition}, pages 1--9, 2015.

\bibitem{Chollet2017}
F~Chollet.
\newblock Xception: Deep learning with depthwise separable convolutions.
\newblock In {\em 2017 IEEE Conference on Computer Vision and Pattern Recognition (CVPR)}, 2017.

\bibitem{Dosovitskiy2020}
A~Dosovitskiy, L~Beyer, A~Kolesnikov, D~Weissenborn, X~Zhai, T~Unterthiner, and et~al.
\newblock An image is worth 16x16 words: Transformers for image recognition at scale.
\newblock {\em arXiv preprint arXiv:2010.11929}, 2020.

\bibitem{MG2001}
G~Mein, G~Neijenhuis, W~Morgan, D~Reinemann, J~Hillerton, J~Baines, I~Ohnstad, M~Rasmussen, L~Timms, J~Britt, and et~al.
\newblock Evaluation of bovine teat condition in commercial dairy herds: 1. non-infectious factors.
\newblock 2001.

\bibitem{MW2023}
M~Wang and P~Lin.
\newblock Supervised learning model for key frame identification from cow teat videos.
\newblock {\em arXiv preprint arXiv:2409.18797}, 2023.

\bibitem{zhangdairy2022}
Y~Zhang, P~Basran, I~Porter, and M~Wieland.
\newblock Dairy cows teat-end condition classification using separable transductive learning.
\newblock In {\em 61st National Mastitis Council (NMC) Annual Meeting}, 2022.

\bibitem{zhang2022separable}
Y~Zhang, I~Porter nad M~Wieland, and P~Basran.
\newblock Separable confident transductive learning for dairy cows teat-end condition classification.
\newblock {\em Animals}, 12(7):886, 2022.

\end{thebibliography}
}

\end{document}